\documentclass{elsart}

\usepackage{graphicx}

\usepackage{amssymb}
\usepackage{amsmath}

\DeclareFontShape{OT1}{cmtt}{bx}{n}{
  <5><6><7><8><9><10><10.95><12><14.4><17.28><20.74><24.88>cmttb10}{}

\renewcommand{\vec}[1]{\pmb{#1}}
\def\d{\delta}
\def\e{\epsilon}

\begin{document}

\begin{frontmatter}


\title{Generating heavy particles with energy and momentum conservation}
\thanks[label1]{Supported in parts by ITMS:26220120007 (Slovakia), MSM~6840770039, and  
LC~07048 (Czech Republic).}
\author{Michal Mere\v{s}}
\address{Univerzita~Mateja~Bela, Tajovsk\'eho 40, 97401~Bansk\'a~Bystrica, Slovakia}
\address{FMFI, Univerzita Komensk\'eho, Mlynsk\'a Dolina, 84248 Bratislava, Slovakia}
\author{Ivan Melo}
\address{\v{Z}ilinsk\'a Univerzita, Univerzitn\'a 1, 01026 \v{Z}ilina, Slovakia}
\author{Boris Tom\'a\v{s}ik}
\address{Univerzita~Mateja~Bela, Tajovsk\'eho~40, 97401~Bansk\'a~Bystrica, Slovakia}
\address{FNSPE, Czech~Technical~University in Prague, B\v{r}ehov\'a~7, 11519~Prague,
Czech~Republic}
\author{Vladim\'ir Balek}
\address{FMFI, Univerzita Komensk\'eho, Mlynsk\'a Dolina, 84248 Bratislava, Slovakia}
\author{Vladim\'ir \v{C}ern\'y}
\address{FMFI, Univerzita Komensk\'eho, Mlynsk\'a Dolina, 84248 Bratislava, Slovakia}





\begin{abstract}
We propose a novel algorithm, called REGGAE, for the generation of momenta of a given sample 
of particle masses, evenly distributed in Lorentz invariant 
phase space and obeying  energy and momentum conservation. In comparison
to other existing algorithms, REGGAE is designed for the use in multiparticle production 
in hadronic and nuclear collisions where many hadrons are produced and a large part 
of the available energy is stored in the form of their masses. The algorithm uses a
loop simulating multiple collisions which lead to production of configurations with 
reasonably large weights. 
\end{abstract}

\end{frontmatter}

\section*{Program summary}

\noindent
\textit{Program title:}  REGGAE (REscatterig-after-Genbod GenerAtor of Events)\\
\textit{Catalogue identifier:}\\
\textit{Program summary URL:} http://www.fpv.umb.sk/{\symbol{126}}tomasik/reggae\\
\textit{Program obtainable from:} http://www.fpv.umb.sk/\symbol{126}tomasik/reggae\\
\textit{RAM required to execute with typical data:}  This depends on the number of particles which are generated. For 10 particles like in the attached example it requires about 120~kB.\\
\textit{Number of processors used:} 1\\
\textit{Computer(s) for which the program has been designed:} PC Pentium 4, though no 
particular tuning for this machine was performed.\\
\textit{Operating system(s) for which the program has been designed:} 
Originally designed on Linux PC with g++, but it has  been compiled and ran successfully on 
OS X and MS Windows with  Microsoft 
Visual C++ 2008 Express Edition, as well.\\
\textit{Programming language:} C++\\
\textit{Size of the package:}  12~KB\\
\textit{Distribution format:}  zipped archive   \\
\textit{Number of lines in distributed program, including test data etc.:}   1468\\
\textit{Number of bytes in distributed program, including test data etc.:}  52~KB\\
\textit{Nature of physical problem:}
The task is to generate momenta of a sample of particles with given masses which obey energy and momentum conservation. Generated samples should be evenly distributed in the available Lorentz invariant phase space. 
\\
\textit{Method of solving the problem:}
In general, the algorithm works in two steps. First, all momenta are generated with the GENBOD algorithm. There, particle production is modelled as a sequence of two-body decays of heavy resonances. After all momenta are generated this way, they are reshuffled. Each particle 
undergoes a collision
with some other partner such that in the pair centre of mass system the new directions of momenta are distributed isotropically. After each particle collides only a few times, 
the momenta are distributed evenly across the whole available phase space. Starting with GENBOD is not essential for the procedure but it improves the 
performance. 
\\
\textit{Typical running time:} This depends on the number of particles and number of events one wants to generate. On a LINUX PC with 2~GHz processor, generation of 1000 events with 10 particles each takes about 3~s. 

\section{Introduction}
\label{s:intro}

A frequent task encountered in simulations of particle production is 
to generate a sample of many particles  which 
exactly conserve energy and
momentum. The events generated in the procedure should be uniformly distributed in the Lorentz 
invariant phase space (LIPS).  
This is a standard textbook problem in the case of two or three-body emission,
but becomes increasingly involved as the number of particles grows. Such situations may also 
become relevant now with the high multiplicity data from proton-proton as well as 
Pb+Pb collisions at the LHC. Algorithms 
have been developed based on the formulas for integration of the LIPS  which 
can be used for such a task. Among them we find GENBOD based on the treatment 
by sequential decay \cite{kopylov60,james68}. 

Originally, the motivation for the development of these algorithms was in the need to 
simulate the Lorentz invariant phase space for Monte Carlo integration of complicated matrix elements. 
The problem of simple algorithms like GENBOD is that for a large number of particles it
generates too often configurations which are very unlikely and have tiny weight of their  
contribution to the phase space integral. If the algorithm is used for event generation, 
such configurations are accepted with very low probability and this makes the simulation 
rather inefficient. 

Improved algorithms are available on the market like RAMBO \cite{Kleiss:1985gy} 
and NUPHAZ \cite{nuphaz}. Those, however, are mostly aimed for the use in 
high energy processes with rather large amount of energy in the form of 
kinetic energy, i.e.\ the masses  small in comparison to the energy. The 
effectiveness drops fast  if the amount of energy contained in the masses grows 
considerably. This is a serious drawback which makes these algorithms 
unusable for the simulation of multiparticle production in nuclear collisions
where masses are of the order of, or larger than, the momenta. 
Also, high multiplicity proton-proton data are beyond reasonable applicability 
of these tools.

When working in this regime a different kind of approach is sometimes used. 
One rather starts with generating the particles from thermal distribution. 
The momentum of the last particle is then calculated from momentum 
conservation and the energies of all particles are rescaled so as to match the 
total energy \cite{toneev}. Another possibility is to calculate the energies and 
momenta of the last pair of particles  \cite{ferroni}. These algorithms are not guaranteed to 
always work and one may have to repeat them in order to generate a usable 
configuration. Moreover, the particles with the calculated momenta may lie far out
off the region where other thermal ones are concentrated. The chance of 
the appearance of such problems grows with the number of particles.

Recently, a new generator was reported in \cite{MCSTHAR} but no details are available
to us about its construction and performance. 

In our approach we first generate $n$-body events which do 
conserve energy and momentum but are not guaranteed to fill the Lorentz invariant phase space 
uniformly.  Then, the generated momenta are reshuffled in (virtual) two-body
collisions. A few collisions per particle  lead the 
system to the most likely configuration. In 
this way we solve the problem of generating unlikely configurations: our 
algorithm samples the space of all possible configurations according to their
probability of appearance. This leads to uniform distribution of events within 
the available LIPS. 

Details of the algorithm are explained in 
Section~\ref{s:alg}. This is followed in Section~\ref{s:entropy} by a demonstration
of the approach of events generated by REGGAE to equilibrium after increasing
the number of collisions. To this end we estimate the information 
entropy of the generated configuration. In Section~\ref{s:comp} we compare the
results of REGGAE with RAMBO, NUPHAZ, and the original GENBOD algorithm. 
A short manual for the C++ routines which accompany the paper is 
given in Section~\ref{s:man}.


\section{REGGAE: the algorithm}
\label{s:alg}

We must first generate  hadron momenta so that they will conserve the fixed total 
momentum and energy. Any procedure fulfilling this simple requirement can be used, 
because the final configuration of momenta is formed later in the rescattering part. 
A good choice of the initial generator can, however, make the algorithm more efficient, 
as less improvement is required from the subsequent rescattering part. 
Our procedure starts with GENBOD \cite{james68}. 

We want to distribute energy 
$E$ and momentum $\pmb{P}$ among $n$ particles with masses 
$m_1,\, m_2,\dots m_n$. We can boost into the frame where $\vec{P}$ 
vanishes  and the energy assumes the value $E^*$. After the generation is complete
we boost back to the original frame. The trick is to formally treat the multiparticle 
generation as a sequence of two-body resonance decays. First, an auxilliary resonance
with mass $E^*$ decays into a particle with mass $m_n$ and a resonance with 
mass $M_{n-1}$. Then $M_{n-1}$ goes into $m_{n-1}$ and $M_{n-2}$ etc. 
Directions of the momenta of decay products are  random up to 
momentum conservation. The essence
of the procedure  is in determining the masses of auxiliary resonances. They are 
chosen randomly but they must fulfill the inequalities
\begin{equation}
\label{e:Mineq}
m_i + M_{i-1} \le M_i \le M_{i+1} - m_{i+1}\, ,
\end{equation}
where $M_1 = m_1$ and $M_n = E^*$. The condition is fulfilled if we generate $M_i$'s 
via
\begin{equation}
\label{e:Mgen}
M_i = \sum_{k=1}^{i} m_k + x_i \left ( E^* - \sum_{l=1}^{n} m_l \right )\, ,
\end{equation}
where $x_i$ is random variable from the interval $(0,1)$. Condition (\ref{e:Mineq})
is satisfied if we choose $x_i$'s so that 
\begin{equation}
\label{e:xinq}
x_2 \le x_3 \le x_4 \le \dots \le x_{n-1}\, .
\end{equation}
Details of the procedure can be found in \cite{james68}.

As we explained in the introduction, momentum configurations generated by 
GENBOD often come out with very small weight. In practice, this means that such an event 
is extremely unlikely to appear in Nature. Therefore, in the next step we 
reshuffle the momenta
in order to achieve a more likely configuration. 

In the routine, the momenta of all particles are stored in an array. We randomly 
choose a distance $d$ between positions within the array. Then each particle collides 
with a partner that is stored by 
$d$ positions further down the array. Particles in the last $d$ positions collide with partners in the beginning of the array. In every turn each particle collides twice. 

The collisions are simple $s$-wave scatterings. We always Lorentz boost into the center of mass
system of the colliding pair, generate new directions of the momenta with isotropic distribution,
and then boost back to the original frame. For a small 
number of particles 6 collisions per particle is enough to reach a distribution with 
large weight. This number becomes larger for large numbers of particles.


\section{The approach to most likely configurations}
\label{s:entropy}


For very large number of rescatterings our procedure leads to
uniform filling of the available Lorentz invariant phase space. 
The first step to prove this is to construct the distribution of outgoing
particles in two-particle collisions. 
The distribution is defined as the probability $dP$ of
finding the outgoing momenta $({\vec p}_1, {\vec p}_2)$
in a given elementary
volume $d^3 p_1 d^3 p_2$ of the two-particle momentum space, given the
incoming momenta $({\vec q}_1, {\vec q}_2)$. If the
distribution is isotropic in the center of mass system, we
have in a laboratory system
\begin{equation}
dP = \frac {1}{2\pi} \left(1 - \frac{2M^2}{q^2} + 
\frac{\mu^4}{q^4} \right)^{-1/2} \d^4 (p - q) 
\frac {d^3 p_1 d^3 p_2}{E_1 E_2}, \label{eq:P}
\end{equation}
where $q^\mu$ and $p^\mu$ are the total four-momenta of incoming and
outgoing particles, respectively, $(E_1, E_2)$ are the energies of outgoing
particles, and $M^2$ and $\mu^2$ is the sum and the difference of
the masses of particles squared, $M^2 = m_1^2 + m_2^2$ and
$\mu^2 = m_1^2 - m_2^2$. The crucial observation is that the
distribution in (\ref{eq:P}) normalizes not only in outgoing
momenta but also in incoming ones, i.e.\ it satisfies
\begin{equation}
\int \frac 1{2\pi} \left(1 - \frac {2M^2}{q^2} + \frac
{\mu^4}{q^4} \right)^{-1/2} \d^4 (p - q) \frac {d^3 q_1 d^3
q_2}{\e_1 \e_2} = 1, \label{eq:norm}
\end{equation}
where $(\e_1, \e_2)$ are the energies of incoming particles. This
is seen immediately after replacing $q^2$ in the expression in
front of $\d$-function by $p^2$, since (\ref{eq:norm}) then
reduces to the normalization in outgoing momenta with $({\vec p}_1,
{\vec p}_2)$ renamed to $({\vec q}_1, {\vec q}_2)$ and {\it vice
versa}. Thanks to (\ref{eq:norm}), collisions with the
distribution of outgoing particles (\ref{eq:P}) produce an
equilibrium distribution of the system of $n$ particles that is
uniform in LIPS
\begin{equation}
dp = C_P \d^4 (p_{\mathrm{tot}} - P) \prod_{i=1}^n \frac {d^3 p_i}{E_i},
\label{eq:p1}
\end{equation}
where $p_{\mathrm{tot}}^\mu$ is the total four-momentum of the system, $P^\mu$
is the value assigned to $p^\mu_{\mathrm{tot}}$ and $C_P$ is a normalization
constant depending only on $P^\mu$. Indeed, if we view $dP$ for an
arbitrary pair of particles as a two-particle block of a
block-diagonal $n$-particle transfer matrix, it follows from
(\ref{eq:norm}) that (\ref{eq:p1}) is an eigenvector of the
transfer matrix corresponding to the eigenvalue 1. Then, 
Perron-Frobenius theorem guarantees that it is the \emph{only}
eigenvector with that property. (The last claim holds if the
particle momenta can change from any values to any other values
consistent with the conservation of energy-momentum in a finite
number of collisions, which seems plausible enough.) 
The state of the system described by the eigenvector of the 
transfer matrix with the eigenvalue 1 is obviously an equilibrium 
state, since it does not change in the evolution of the system.
Another consequence of the Perron-Frobenius theorem is that the 
system evolves towards it from any nonequilibrium initial state.




While this proves that our procedure will eventually lead to the uniform filling of
the available LIPS, we also want to demonstrate how {\it fast} this happens, {\it i.e.},
how fast the system approaches the most likely (equilibrium)
configurations when starting from unlikely ones generated by GENBOD. In algorithms
based on Monte Carlo integration of the phase space usually a weight is determined for
each generated configuration. This weight could be used as a measure of likelihood of a given configuration,
however, it depends on integration variables and it is not clear that we can define it uniquely in our case. 

We thus use a different approach, with information entropy as a measure of likelihood, defined
not for a single configuration, but for the whole
set of $N_e$ configurations (events), each with $n$ particles. 
We assume that the momenta of all $n N_e$ particles are distributed according 
to the same underlying single-particle probability density distribution $\rho(\vec{p})$. 
The \emph{information entropy} is defined as
\begin{equation}
\label{e:Sdef}
S = - \int_{\Sigma} \rho(\vec{p})\, \ln \rho(\vec{p})\, d^3\vec{p}
\end{equation}
where the integration runs through all accessible momentum space $\Sigma$. 

We want to show that the rescattering part 
of the algorithm changes this entropy in such a way that it will grow from some value corresponding to
unlikely configurations to higher values for more likely configurations and eventually it will 
saturate at equilibrium.
As can be seen, we formulate this calculation for a distribution on a three-dimensional
momentum space.  The distribution $\rho(\vec{p})$ is generally unknown. 
We estimate it from the sample of momenta from the set of  $N_e$ configurations:
first the available three-momentum space is divided into $N$ elementary cells of equal volume $\Delta V_i$,
 then the probability density $\rho(\vec{p})$ in a cell $\Delta V_i$ is estimated 
by $\rho_i$ which is given as
\begin{equation}
\label{e:rho_estimate}
\rho_i = \frac{n_i}{n N_e \Delta V_i} \, ,
\end{equation}
where $n_i$ is the number of particles which fall in the cell $\Delta V_i$.

Then the entropy is estimated as
\begin{equation}
\label{e:S_estimate}
S = - \sum_{i=1}^N \left ( \rho_i \ln{\rho_i}\right ) \Delta V_i \, ,
\end{equation} 
where $N$ is large and $\Delta V_i \sim 1/N$ small.

At equilibrium the information entropy can be calculated theoretically in the limit of large $n$,
since $\rho(\vec{p})$ is known in this case. It is related to 
the single-particle energy distribution $\rho_E(E)$ (called here the LIPS-Boltzmann distribution),
which  is
derived from the uniform distribution in LIPS using the 
Darwin-Fowler method and is given by \cite{e:Young}
\begin{equation}
\label{e:LIPS-Boltzmann1}
\rho_E(E) dE = {\cal N} \sqrt{E^2 - m^2} \exp\left (-\frac{E}{k_B T} \right ) dE ,
\end{equation} 
where $E = \sqrt{p^2+m^2}$ is the total energy of a single particle, 
$T$ is the LIPS-temperature, $m$ is the mass of the particle and $k_B$ 
the Boltzmann's constant.
\[
{\cal N} = \left ( m k_B T \mbox{K}_1 \left ( \frac{m}{k_B T} \right ) \right )^{-1}
\]
is a normalization constant, and $\mbox{K}_1(m/k_B T)$ is the modified Bessel function.
Note that $\rho_E(E)$ 
differs from the canonical Boltzmann distribution in two important aspects: the canonical distribution has an extra $E$ in front of the exponential and the canonical temperature is different from the LIPS-temperature $T$. The differences come essentially from the fact that unlike the canonical relativistic Boltzmann gas which thermalizes via collisions both in configuration and momentum space, our system thermalizes via collisions which take place only in the momentum space. The LIPS-temperature $T$ is implicitly defined by the equation for the mean value of the single-particle energy
\begin{equation}
\label{e:LIPS-T}
\left \langle E \right \rangle  = \int_m^{\infty} E \rho_E(E) dE  = 
m \frac{\mbox{K}_2(m/k_B T)}{\mbox{K}_1(m/k_B T)} = \frac{E^*}{n}\,  .
\end{equation}  
Then, the equilibrium probability density $\rho(\vec{p})$ is found from the LIPS-Boltzmann distribution
\begin{equation}
\label{e:LIPS-Boltzmann2}
\rho(\vec{p}) = \frac{\rho_E(E) dE}{d^3\vec p} =  \frac{\rho(E) dE}{4 \pi p^2 dp} = \frac{N}{4 \pi} \frac{1}{\sqrt{p^2 + m^2}} \exp{\left (-\frac{\sqrt{p^2 + m^2}}{k_BT} \right )}\, .
\end{equation}

\begin{figure}
\includegraphics[width=1.0\textwidth]{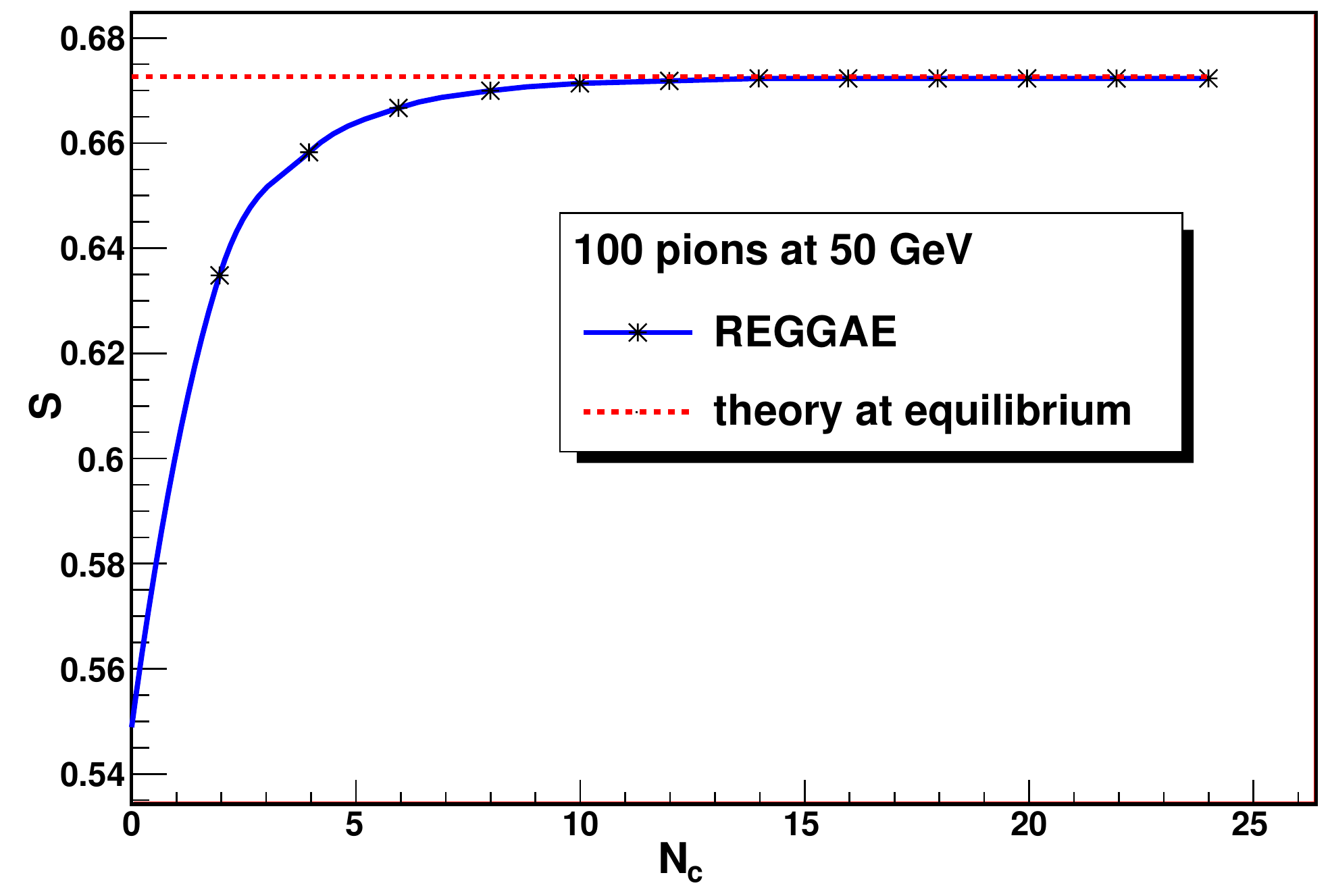}
\caption{Evolution of the information entropy estimates with the number of collisions $N_c$ for $10^5$ REGGAE-generated configurations of $n=100$ pions ($m=139$ MeV) of the total energy $E^*=50$ GeV. The red line shows the 
theoretical equilibrium value $S=0.6724$, the blue curve shows the estimates of Eq. \ref{e:S_estimate} based on REGGAE configurations. The statistical errors of the estimates $\sigma_S < 0.0001$.   
\label{f:entevol}}
\end{figure}

The equilibrium information entropy is then found by substituting $\rho(\vec{p})$ into the definition of $S$ in 
Eq. (\ref{e:Sdef}) and integrating. For configurations of $n=100$ pions ($m=139$ MeV) of the total energy $E^*=50$ GeV we get the LIPS-temperature $k_B T=0.20688$ GeV and the equilibrium information entropy prediction $S=0.6724$.
Our expectation is that the entropy estimated from Eq.~(\ref{e:S_estimate}) will grow with the number of collisions $N_c$ and for some $N_c$ it will saturate at the equilibrium value which should coincide with the theoretical prediction.
We show the result of this calculation in Figure~\ref{f:entevol}.

The entropy saturates at the predicted equilibrium value after 12 colisions in this particular case.
For a smaller $n$, we found that the entropy saturates after a smaller number of collisions. 
Our findings are in line with studies of kinetic theory. For example, in \cite{flp} it has been shown that a few collisions are enough to drive the system towards equilibrium distribution.

This indicates that our algorithm can generate relevant momentum configurations not only {\it in principle}, but it can do so in a reasonably small number of steps. Further support that this is indeed so will come from the next section where we apply REGGAE to Monte Carlo integration.

Note here also, that the rescattering algorithm would arrange the momenta
correctly even if we did not start with GENBOD. We have checked this by
starting with 15 equal momenta all in the same direction and another 15 ones
of equal size and opposite direction. Rescattering succeeded in arranging the
momenta but needed more steps to reach saturation. This shows that starting
with GENBOD is not essential, but improves the effectivness. Calculations
using GENBOD were indeed faster.



\section{Comparison with other algorithms}
\label{s:comp}

One might argue that the information entropy saturation at the equilibrium value demonstrated in the previous section
only shows that
the {\it single-particle} momentum distribution has LIPS-thermalized while it is not clear that {\it many-particle} 
momentum distributions have done the same. To provide further support for a fast, few-step LIPS-thermalization,
we applied our algorithm to Monte Carlo integration over the phase space, a procedure sensitive to
many-particle distributions. In this section we also compare REGGAE with other routines available on the market. 

Generally, the infinitesimal 
element of the LIPS for $n$ particles is 
\begin{equation}
\label{e:phi}
d\Phi_{n}  =  \frac{d^3{\vec p_1}}{(2\pi)^3 2 E_1}\frac{d^3{\vec p_2}}{(2\pi)^3 2 E_2}
\dots \frac{d^3{\vec p_n}}{(2\pi)^3 2 E_n} \delta^4(P-p_1-p_2-p_3 \dots -p_n) 
\end{equation}
The general task is to calculate integral over available phase space $\Phi_n$
\begin{equation}
I = \int_{\Phi_n} f(\{ p_i \} )\, d\Phi_n\, , \qquad \{p_i\} = (p_1,p_2,p_3,\dots, p_n)
\end{equation}
where $f(\{ p_i \} )$ is some function of the momenta. Algorithms for Monte Carlo 
integration generate configurations of momenta, calculate their weights $w_j$ 
and then approximate the integral  with (sample mean method) 
\begin{equation}
\label{e:Iapp2}
\hat I = \left ( \frac{\sum_{j=1}^{N} w'_j f(\{p_i\}_j )}{\sum_{j=1}^{N} w'_j} \right )\, \Phi_n\, ,
\end{equation}
where $w'_j = w_j/w_{max}$ is the ratio of the weight of the configuration to the maximum weight calculated 
or found empirically in the large set of configurations and
the sum runs over $N$ different momentum configurations 
$\{p_i\}$.

If the configurations correspond to real events, 
they have $w'_j = 1$ and
\begin{equation}
\label{e:Iapp}
\hat I = \left ( \frac{1}{N} \sum_{j=1}^{N} f(\{p_i\}_j ) \right )\, \Phi_n\, 
\end{equation}
We have tested the sum of Eq.~(\ref{e:Iapp}) calculated  with events generated 
by our algorithm against calculations by 
RAMBO \cite{Kleiss:1985gy}, NUPHAZ \cite{nuphaz}, and GENBOD \cite{james68}.
In case of GENBOD we accepted the generated events with the probability determined 
by the ratio of the event weight and the maximum theoretical event weight.
This we call weighted GENBOD (wGENBOD). It should be distinguished from 
the unweighted GENBOD (uGENBOD) where we ignore the weights, which is incorrect and 
we show it here for the sake of demonstration. We have performed the tests on a variety 
of functions. Here we present just one example.

We integrated a function of five four-momenta
\begin{equation}
\label{e:f5}
f_5(p_1,p_2,p_3,p_4,p_5) = 
\frac{(p_1^2 + p_2^2 + p_3^2)p_1^2}{M^4 + p_4^2 p_5^2}
\end{equation}
with $M^4 = 25\, \mbox{GeV}^4$. The phase space was given by the total four-momentum 
$P = (100,0,0,0)\, \mbox{GeV}$ which was distributed among particles with the mass of
$m = 1$~GeV. 

In Table~\ref{t:iT3}
\begin{table}[t]
\caption{%
Mean values of $f_5$ from eq.~(\ref{e:f5}) calculated with five different generators.
Number of momentum configurations (events) used in the calculations is $N$. Each event consists of
$n = 5$ particles with $m = 1$~GeV.
The last row shows the time for computation with $10^7$ configurations. 
\label{t:iT3}}
\begin{center}
\begin{tabular}{|c|c|c|c|c|c|}
\hline
N & REGGAE ($N_c = 6$) & NUPHAZ & RAMBO & wGENBOD & uGENBOD\\
\hline 
$10^4$ & 195.4 & 270.48 & 181.60 & 205.58 & 1512 \\
\hline 
$10^5$ & 214.5 & 223.62 & 199.24 & 203.81 & 1610 \\
\hline
$10^6$ & 211.2 & 212.56 & 211.63 & 208.72 & 1635 \\
\hline
$10^7$ & 209.8 & 207.71 & 208.97 & 209.334 & 1655 \\
\hline
time   & 26 min & 5 min & 1 min & 7 min & 7 min  \\
\hline
\end{tabular}
\end{center}
\end{table}
we compare the results of the five generators together with the times of computation 
of $10^7$ events on AMD Athlon(tm) 64 X2 Dual Core Processor  
(speed: 3 GHz, RAM: 2 GB,
OS: Ubuntu Linux, kernel: 2.6.31-22).
We show the mean values of $f_5$, i.e.\ only the part of  formula~(\ref{e:Iapp}) within the parentheses.
The algorithms, including REGGAE, yield consistent results except for the unweighted 
GENBOD, which was expected. 
Although the result of REGGAE is fine, its computation time seems discouraging. 
This changes if we increase the number of particles with  the same mass while 
keeping the same total energy, see Table~\ref{t:iT4}. We also show the dependence of the
result on the number of collisions $N_c$ which 
the particles suffer in REGGAE. The $N_c = 6$ case 
starts to differ slightly from the other algorithms which can 
be fixed by increasing $N_c$ to $12$. 
\begin{table}[t]
\caption{Mean values of  $f_5$ as in Table~\ref{t:iT3}, but for $n = 30$ particles with $m = 1$~GeV. The dependence on the number of collisions in REGGAE is also shown. The last row shows the time for computation with $10^6$ configurations.
\label{t:iT4}}
\begin{center}
\begin{tabular}{|c|c|c|c|c|c|c|}
\hline
N          & REGGAE    & REGGAE    & REGGAE    & NUPHAZ & RAMBO & wGENBOD \\
           & $N_c = 6$ & $N_c = 8$ & $N_c = 12$ &        &       &         \\
\hline 
$10^4$     & 13.63     & 13.41     & 13.78     & 12.77  & 13.23 & 12.23   \\
\hline 
$10^5$     & 13.99     & 13.52     & 13.36     & 13.15  & 13.21 & -       \\
\hline
$10^6$     & 13.84     & 13.42     & 13.19     & 13.06  & 13.12 & -       \\
\hline
time       & 16 min  & 20 min    & 28 min    & 6 min  & 11 min & 300 min  \\
\hline
\end{tabular}
\end{center}
\end{table}
Our algorithm is clearly most effective in cases where the total mass 
of particles makes up a large part of the energy budget in the simulation,
see Table~\ref{t:iadd} where the total mass of particles $n m = 60$ GeV represents $60 \%$ of the
energy budget (RAMBO and GENBOD were too slow to be included in the Table).
\begin{table}[t]
\caption{Mean values of $f_5$ as in Table~\ref{t:iT3}, but for $n = 60$ particles with $m = 1$~GeV. The last row shows the time for computation with $10^5$ configurations.
\label{t:iadd}}
\begin{center}
\begin{tabular}{|c|c|c|}
\hline
N          & REGGAE   ($N_c = 12$)  & NUPHAZ  \\
\hline 
$10^4$         & 0.6339 & 0.6185  \\
\hline 
$10^5$         & 0.6185 & 0.6315  \\
\hline
time for $10^5$ & 6 min & 63 min \\
\hline
\end{tabular}
\end{center}
\end{table}
This is not surprising: we developped REGGAE for the use 
in simulations of multiparticle production in nuclear and high multiplicity 
hadronic collisions 
while the other methods were derived for applications in high energy physics
where most of the available energy often goes into momenta. 

For large $n$, REGGAE can be the fastest even in the case when $90 \%$
of the available energy goes into momenta and just $10 \%$ is in the particle masses. 
The time to generate $10^4$ configurations for $n = 500$ particles with the mass 0.1~GeV and the total four-momentum 
$P = (500,0,0,0)\, \mbox{GeV}/c$ is 8 min for REGGAE ($N_c = 10$) and 23 min for NUPHAZ.


\section{REGGAE: a reference manual}
\label{s:man}

The package comes with the following files: 
\begin{description}
\item{\verb=Makefile=} for linux-like systems with g++ compiler this is the simple Makefile to compile the example program. 
\item{\verb=example.cpp=}  is the file with the main routine used to illustrate the action of REGGAE. You do not need this file if you want to embed REGGAE in your simulation.
\item{\verb=example.dat=} is the output file of the code if run in the form as it is distributed.
\item{\verb=reggae.cpp=} is the file where main algorithms are implemented.
\item{\verb=reggae.hpp=} the header file which must be included in the application that is supposed to use REGGAE.
\item{\verb=specrel.cpp=} supporting file with definitions of classes for four-vectors, Lo\-rentz boosts etc.
\item{\verb=specrel.hpp=} header file for  \verb=specrel.cpp=.
\end{description}

To use the method, the user must define the following variables: 
\begin{description}
\item{\verb=int n=} the number of particles
\item{\verb=double * amass=} which must be allocated to include \verb=n= values for the masses of the particles. Before calling the routine the array must be filled with the masses. 
\item{\verb=vector4 * avec=} which must be allocated to hold \verb=n= four-vectors. This array will hold the particle four-momenta. 
\item{\verb=vector4 P=} the total four-momentum. Before calling the routines it must posses a value. 
\item{\verb=long int seed=} is the initial seed for the random number generator. It may be initialised always with the same value or with the help of the machine time. 
\end{description}
The four-vector class \verb=vector4= is defined in \verb=specrel.hpp=. It has been designed so that its use is intuitive. To get or set the component of the four-vector use [*] (e.g. avec[0], numbering runs
from 0 to 3). Minkowski metric (+,--,--,--) is applied so that \verb=a*b= gives the proper 
four-vector product. 

To generate a sample of momenta in REGGAE, first the GENBOD algorithm  must be called by 
\begin{verbatim}
Mconserv(P,n,amass,avec,&seed);
\end{verbatim}
where the variables have been explained above. Now the momenta are stored in \verb=avec=. The second step is to call 
\begin{verbatim}
collision(n,avec,&seed);
\end{verbatim}
This step reshuffles momenta and returns them also in the array \verb=avec=. This array of momenta is the result of the generator. 

The procedure is illustrated in the main routine distributed with the package. 

By default, \verb=collisions= is set to run eight scatterings per particle. It is possible to decrease this number in order to run faster or increase it in order to have more confidence that the generated configurations are in the regime of saturated entropy. This is done by setting the variable \verb=int RG_opak= to the value equal to half of the number of collisions.


\section{Summary}
\label{s:summ}

REGGAE gives the possibility of robust and effective simulation of events 
with few as well as many particles with  strict conservation of momentum and energy. This is very much 
needed feature in many models used for the simulation of multiparticle production.


\paragraph*{Acknowledgements}
We thank Evgeni Kolomeitsev for stimulating discussions.
This work was partially supported by the Agency of the Slovak
Ministry of Education
for the Structural Funds of the EU, under project ITMS:26220120007.
BT gratefully acknowledges financial support via grants No.\ MSM~6840770039, and  
LC~07048 (Czech Republic). MM and BT thank the CERN PH-TH department  where a part of this work was completed for the kind hospitality. 


\end{document}